\documentstyle[12pt]{article}
\def\lsim{\hbox{$\hskip0.3em\raisebox{-.4ex}{$\sim$}\raisebox{.4ex}{\hskip-0.8em$<$\hskip0.1em}$}}

\newcommand{\NN}{\nonumber}
\newcommand{\quadtwo}{\quad\quad}

\begin{document}
%%%%%%%%%%%%%%%%%%%%%%%%%%%%%%%%%%%
%%%%%%%%%   Cover Page$B!!!!(B%%%%%%%%%
%%%%%%%%%%%%%%%%%%%%%%%%%%%%%%%%%%%
\title{
\begin{flushright}
{\normalsize YITP-K-1053}
\end{flushright}
\vspace{1cm}
Finite Size Effects and Conformal Symmetry of $O(N)$ Nonlinear 
$\sigma$ Model in Three Dimensions}
\author{Akira FUJII
\thanks{e-mail address: {\tt fujii@jpnyitp.bitnet}} 
and Takeo INAMI
\\ \\ 
{\sl Yukawa Institute for Theoretical Physics,} \\ 
{\sl Kyoto University,} \\ 
{\sl Kyoto 606-01, Japan.}}
\maketitle
\begin{abstract}
We study the $O(N)$ nonlinear $\sigma$ model on a three-dimensional 
compact space $S^1 \times S^2$ (of radii $L$ and $R$ respectively) 
by means of large $N$ expansion, 
focusing on the finite size effects and conformal symmetry of this 
model at the critical point. We evaluate the correlation length and
the Casimir 
energy of this model and study their dependence on $L$ and $R$. We
examine the modular transformation properties of the partition
function, and study the dependence of the specific heat on the
mass gap in view of possible extension of the
$C-$theorem to three dimensions.
\end{abstract}
\newpage
\baselineskip=24pt
%%%%%%%%%%%%%%%%%%%%%%%%%%%%%%%%%%%%%%
%%%%%%%      INTRODUCTION       %%%%%%
%%%%%%%%%%%%%%%%%%%%%%%%%%%%%%%%%%%%%%
\section{Introduction}
In two-dimensional field theories and statistical systems the conformal 
invariance provides a powerful tool which allows us 
to compute exactly physical quantities such as
 correlation functions, critical exponents and 
field contents \cite{BPZ}. This is due to the fact that the conformal 
symmetry is infinite-dimensional in two dimensions. Furthermore, 
we can classify conformal field theory (CFT) models and find their 
field contents 
by studying the finite-size effects and the modular transformation 
properties of such theories 
on the torus. 
Above two dimensions 
the conformal symmetry is finite-dimensional, 
hence this symmetry appears less useful in analyzing 
critical systems. 
Despite this difficulty, 
extension of some of the notions of the two-dimensional 
CFT to higher-dimensional theories has 
been pursued by a few authors \cite{polyakov}. 
In particular, Cardy has studied modular transformation 
properties of the partition 
function of conformal invariant free field theories on a compact 
space of three and higher 
dimensions $D$ \cite{cardy2,cardy1,cardy3}.
He has shown that the partition function of such theories 
on $S^1 \times S^{D-1}$ of radii $L$ and $R$ respectively and 
a moduli $\delta=L/R$, or its derivative with respect to $\delta$, 
is invariant under $\delta\rightarrow 1/\delta$. 
It is important to examine whether interacting field theories have 
the same modular transformation properties 
at their fixed points where they are conformally 
invariant.

In this paper we will consider the $O(N)$ nonlinear $\sigma$ (NL$\sigma$) 
model on a three-dimensional compact space as an example of models 
including interactions. Of a variety of three-dimensional compact
spaces, $S^{1} \times S^{2}$ allows us to study 
the properties with respect to conformal invariance of the
field theory defined on it \cite{cardy3}. 
In ${\bf R}^{D}$ with $D$ less than four $O(N)$ NL$\sigma$ model is 
known to be renormalizable in the $1/N$ expansion and permits us a 
nonperturbative analysis. Furthermore, 
the model possesses 
order (symmetry breaking)-disorder phase transition in 
dimensions $2<D<4$ \cite{arefeva} 
and hence it serves our purpose of studying the modular properties at the 
critical value of the coupling constant. 
In condensed matter physics, 
the $(2+1)$-dimensional $O(3)$ NL$\sigma$ model 
has been studied as an 
effective field theory of the long wavelength behavior of 2-dimensional 
quantum antiferromagnet \cite{hal}. 
Recently, from the analyses of the large $N$ $O(N)$ 
NL$\sigma$ model on the semi-compact 
space $S^1 \times {\bf R}^{2} $, interesting results 
have been obtained on the low temperature properties of the quantum 
antiferromagnet  \cite{spin}. 
Curiously the specific heat of $O(N)$ NL$\sigma$ model is represented by 
Roger's polylogarithmic functions \cite{sachdev}; this is analogous to 
the specific heat of certain two-dimensional integrable systems 
being represented by Roger's dilogarithmic function \cite{bethe}.

%.
\section{Saddle Point Method and $\beta-$ Functions }
%
%%%%%%%%%%%%%%%%%%%%%%%%%
%%%%%%% Definition%%%%%%%
%%% partition function %%
%%%%%%%%%%%%%%%%%%%%%%%%%
The $O(N)$ NL$\sigma$ model 
in the $D$-dimensional Euclidean space ${\sl M}$ is defined by the action
\begin{equation}
S={1 \over 2g}\int_{\sl M} d^{D}x \partial_{\mu}{\vec n}\cdot 
\partial^{\mu}{\vec n},
\end{equation}
where ${\vec n}$ is an $N$-component vector obeying the constraint 
${\vec n}^2 =1$. By introducing an auxiliary field $\mu$, the partition 
function is expressed as 
\begin{equation}
Z=\int D{\vec n}D\mu
\exp\Bigl(-{1 \over 2g}\int_{\sl M}
d^{D}x \mbox{[} \partial_{\mu}{\vec n}\cdot
\partial^{\mu}{\vec n} + \mu({\vec n}^2 -1) \mbox{]}\Bigr) \label{action3} 
\end{equation}
By carrying out the path integration of ${\vec n}$,
the partition function (\ref{action3}) is written in terms of 
the effective action as follows.
\begin{eqnarray}
Z&=&\int D\mu \exp\Bigl( -(N/2)S_{\it eff}\Bigr) , \label{efpart} \\ 
S_{\it eff}&=&\int_{\sl M}d^D x\Bigl( -{1\over {\tilde g}}\mu
+Tr\ln(-\Delta +\mu)\Bigr) , 
\label{effective}
\end{eqnarray}
where we have set ${\tilde g}=gN$. 

We consider the limit 
$N\rightarrow \infty$ keeping ${\tilde g}$ fixed. In this limit 
we can evaluate 
the partition 
function (\ref{efpart}) by means of the saddle point method. 
The gap equation for the saddle point value of $\mu$ is 
\begin{equation}
{1 \over {\tilde g}}=Tr {1 \over -\Delta +\mu}. \label{saddle}
\end{equation}
It is known that 
the $O(N)$ NL$\sigma$ model on ${\bf R}^3$ has an 
infrared (IR) fixed point at $g=0$  
and an ultraviolet (UV) fixed point at $g=g_{c}$ for $2<D<4$. 
We will consider the model on ${\sl M}=S^1 \times S^2$ (of radii $L$ and
$R$ respectively) and solve the gap equation (\ref{saddle}) 
at the UV fixed point at $g=g_{c}$, 
at which the model is transformed by a conformal mapping into a theory
which possesses conformal invariance in ${\bf R}^{3}$. 

We recapitulate the renormalization group (RG) 
transformation of the $O(N)$ NL$\sigma$ model \cite{arefeva,spin}.
In accordance with the saddle point approximation, we calculate 
the $\beta$-function of the $O(N)$ NL$\sigma$ model in the 
large $N$ limit to obtain the fixed point. 
Since the UV divergence behavior 
is independent of the global property of the space \cite{text}, 
we may consider the model 
on $S^1 \times {\bf R}^{2} $ (of radius $L$) instead of $S^1 \times S^2 $. 
It is useful to introduce the weak constant magnetic field 
${\vec h}$ for the purpose of computing the wave function
renormalization constant. We choose the 
the first component of ${\vec n}$ in the direction of 
${\vec h}=(h,{\vec 0})$ and
set ${\vec n}(x)=(\sigma(x),{\vec \pi}(x))$. 
 The partition function is 
now given, after performing the path integral for $\sigma(x)$, by
\begin{eqnarray}
Z&=&\int {D{\vec \pi}(x)\over \sqrt{1-{\vec \pi}(x)^2 }}
\exp \{ -{1\over 2g}\int^{L}_{0}dx_0 \int d^2 x
((\partial_{\mu}{\vec \pi})^2 
+{({\vec \pi}\partial_{\mu}{\vec \pi})^2 \over 1-{\vec \pi}^2 })+ \nonumber \\ 
&& \ \ \ \ \ \ \ \ \ \ \ \ \ \ \ \ \ \ \ \ \ \ \ \ \ \ \ \ \ \ \ \ \ \ \ \ \ \ 
+h\int^{L}_{0}dx_0 \int d^2 x\sqrt{1-{\vec \pi}^2 } \} .
\label{ps2}
\end{eqnarray}

In parallel with field theories at finite temperature, UV divergences
of the present model can be handled by introducing two bare 
coupling constantsb \cite{text}, 
\begin{equation}
{\tilde g}=gN,\quadtwo t={\tilde g}/L .
\end{equation}
The RG transformation in the theory defined by 
(\ref{ps2}) amounts to changing the momentum 
cut-off in ${\bf R}^2$ from $\Lambda$ to 
$e^{-l}\Lambda$. $a=\Lambda^{-1}$ plays the role of the lattice
spacing in the lattice version. 
The $\beta$-functions for ${\tilde g}$ and 
$t$ can be calculated by carrying out 
this RG transformation in the momentum space. The result is: 
\begin{eqnarray}
\beta_{\tilde g}=-{d{\tilde g}\over dl}&=&{\tilde g}
-{{\tilde g}^2 \over 4\pi}
\Lambda 
\coth{L\Lambda\over 2}, \label{beta1} \\ 
\beta_t =-{dt\over dl}&=&-{{\tilde g}t\over 4\pi} 
\Lambda 
\coth{L\Lambda\over 2}, \label{beta2}
\end{eqnarray}
where we have put $h=0$. We find that the RG transformation has two fixed 
points, 
\begin{equation}
({\tilde g}_c ,t_c )=(0,\mbox{any value})\ \ \ \ {\rm and} \ \ \ \ \ 
(4\pi/\Lambda,0)
,
\end{equation}
where we have set $\coth(L\Lambda/2)=1$ by considering the limit 
$L\Lambda=L/a\rightarrow\infty$. 
The fixed point ${\tilde g}_c =0$ is IR stable and 
${\tilde g}_c =4\pi/\Lambda $ is UV stable.
We note that the critical coupling constants ${\tilde g}_c$ are 
the same as those on ${\bf R}^3 $ as they should be.
The $\beta$-functions (\ref{beta1}) and (\ref{beta2}) 
can also be obtained in the saddle point 
method \cite{nefr}. The $\beta(t)$ vanishes only at $t=0$. This
reflects the fact that no phase transition can occur at finite 
$L$ in the spacetime $S^{1} \times {\bf R}^{2}$; it is a special case of
the Mermin-Wagner-Coleman's theorem \cite{mwc}.

The above result was derived by adopting the periodic boundary condition (PBC) 
in the $S^1 $ coordinate. The $\beta$-functions for the model with the 
antiperiodic boundary condition (APBC) can be derived by repeating the same RG 
transformation as above. The result differs from eqs. 
(\ref{beta1}) and (\ref{beta2}) slightly: 
\begin{eqnarray}
\beta_{\tilde g}&=&{\tilde g}
-{{\tilde g}^2 \over 4\pi}
\Lambda 
\tanh{L\Lambda\over 2}, \label{beta3} \\ 
\beta_t &=&-{{\tilde g}t\over 4\pi} 
\Lambda 
\tanh{L\Lambda\over 2}. \label{betaap}
\end{eqnarray}
The UV stable critical value of ${\tilde g}$ is 
${\tilde g}_{c}=4\pi /\Lambda $ and 
 coincides with that in the case of PBC, 
as it should.
\newline
%%%%%%%%%%%%%%%%%%%%%%%%%%%%%%
%%%%%  The Gap Equation %%%%%%
%%%%%%%%%%%%%%%%%%%%%%%%%%%%%%
\section{$O(N)$ Nonlinear $\sigma$-Model on $S^{1} \times S^{2}$}
 We are now ready to investigate the 
$O(N)$ NL$\sigma$ model on $S^1 \times S^2 $.
We consider this model with the value of ${\tilde g}$ fixed at 
its UV critical value ${\tilde g}_{c}$ 
while the value of $t$ being left arbitrary.  
First we write 
the gap equation (\ref{saddle}) explicitly.
The eigenvalues of the Laplacian $\Delta$ on $S^1 \times S^2 $ are 
$\omega_{n+\epsilon }^2 +l(l+1)/R^{2}$, 
where $\omega_{n+\epsilon /2}=2\pi (n+\epsilon/2)/L$ with 
$\epsilon $ being 0 or 1 depending on whether we take 
the PBC or APBC in the direction of $S^1 $ respectively. 
$l$ takes non-negative integers, and $n$ integers. 
Write $l(l+1)/ R^2 +\mu=(l+1/2)^2 / R^2 +\lambda^2 $, 
where $\lambda^2 =\mu -1/4R^2$. 
The gap equation (\ref{saddle}) takes the form 
\begin{equation}
{1 \over {\tilde g}}={1 \over 32\pi^3 R^2 L}
\sum_{l=0}^{\infty} 
\sum_{n=-\infty}^{\infty} 
{2l+1 \over (n+{\epsilon \over 2})^2 /L^2 +(l+1/2)^2 /4\pi^2 R^2 
+\lambda^2 /4\pi^2 }. \label{sadmode}
\end{equation}
%%%%%%%%%%%%%%%%%%%%%%%%%%%%%%%
%%%%%%The solution on S1*S2 %%%
%%%%%%%%%%%%%%%%%%%%%%%%%%%%%%%
We solve this equation (\ref{sadmode})
by keeping ${\tilde g}$ fixed at its critical value ${\tilde g}_{c}$ 
and find the critical value $\lambda_{c}$ of $\lambda$ or $\mu_c$ of $\mu$. 
%%%%%%%%%%%%%%%%%%%%%%%%%%%%%%%%%%%%%%%%%%%%%%%
%%%%Under Periodic Boundary Condition %%%%%%%%%
%%%%%%%%%%%%%%%%%%%%%%%%%%%%%%%%%%%%%%%%%%%%%%%
%
\subsection{Case of periodic boundary condition (PBC)}\label{s:pbc} 
We solve the gap equation (\ref{sadmode}) in the three limiting 
cases $R=\infty$, $L=\infty$ and $a\ll L\ll R$.
%\renewcommand{\theenumi}{(1\alph{enumi})}
%\begin{enumerate}
%\item 

(1a)\quad $R=\infty $\ \ ($S^1 \times {\bf R}^2 $).

The gap equation (\ref{sadmode}) is further reduced to 
\begin{equation}
{1 \over {\tilde g}}=
{1\over 4\pi L}\ln\prod_{n=-\infty}^{\infty}
{(n^2 +(\Lambda L/2\pi))^2 \over (n^2 +(\lambda L/2\pi))^2}=
{1 \over {\tilde g_c }}-{1\over 2\pi L}\ln (2
\sinh{L\lambda \over 2}). \label{e0a}
\end{equation}
On substituting the critical value 
${\tilde g}={\tilde g}_{c}=4\pi / \Lambda $ 
for ${\tilde g}$ on the left side of (\ref{e0a}), 
we get 
$\sinh (L\lambda_c /2)=1/2$, and hence 
\begin{equation}
\lambda_c ={2\over L}\sinh^{-1}{1\over 2}
={2\over L}\ln{1+\sqrt{5} \over 2}
=0.9624/L.
\end{equation}
Regarding the length of $S^1 $ as the inverse of the temperature, 
we have the $O(N)$ NL$\sigma$-model at finite temperature. 
Then $\xi =\lambda_{c}^{-1}$ is the correlation 
length  at temperature $T=1/L$. 
The specific heat of this model, 
$C=-{\partial\over\partial T}\ln Z$, has been calculated in  \cite{sachdev}
\begin{equation}
{C\over N}=\zeta (3)^{-1}\left(Li_{3}(e^{-L\lambda_{c}})
+(L\lambda_{c})Li_{2}(e^{-L\lambda_{c}})+{(L\lambda_{c})^{3}\over 6}\right)
={4\over 5},
\end{equation}
where $Li_{2}(x)$ and $Li_{3}(x)$ are Roger's 
dilogarithmic and trilogarithmic functions respectively, 
$Li_{p}(x)=\Sigma_{n=1}^{\infty}x^n /n^p $. 
It is curious to see that a rational number emerges 
from polylogarithms. 
It has been observed that Roger's dilogarithmic functions appear in the 
computation of the specific heat of two-dimensional integrable models 
and that a simple rational number comes out by use of the addition
formula
 \cite{yang}. 

%\item  
(1b)\quad $L=\infty $\ \ (${\bf R}\times S^2 $).

The saddle point equation (\ref{sadmode}) can be reduced to 
\begin{equation}
\int {dk\over 2\pi}
\Bigl( {1\over 4\pi R^2 }
\sum_{l=0}^{\infty}{2l+1\over k^2+(l+1/2)^2/4\pi^2 R^2 
+\lambda_{c}^2 /4\pi^2 }
-\int{d^{2}{\bf k}\over (2\pi)^2 }{1\over k^2 +{\bf k}^2} \Bigr) =0.
\end{equation}
We obtain to the order of 
$1/R$ 
\begin{equation}
\sqrt{\mu_c} ={1\over 2R}+O(1/R^3 \Lambda^2 ).
\end{equation}
The anomalous dimension $d$ of ${\vec n}$ 
can be read from the $R$ dependence of the correlation length, 
$\sqrt{\mu_c}=\xi^{-1}=d/R$, 
which follows from finite size scaling \cite{cardy2}. 
We get $d=1/2$, which agrees with the usual computation in the 
large-$N$ limit.

%\item 
(1c)\quad General case.

We consider the case, $a\ll L\ll R$. 
In eq.(\ref{sadmode}), the summation over $l$ is divergent, 
and is regularized by introducing the cutoff $\Lambda\gg 1/L$. 
The summation can be performed by means 
of the Euler-Maclaurin formula 
\begin{equation}
{1 \over N}\sum_{n=n_1}^{n_2}f({n \over N})=
\int ^{n_2 +1/2 \over N}_{n_1 -1/2 \over N}f(x)dx
-{1 \over 24N^2}
\mbox{[}f'({n_2 +1/2 \over N})-f'({n_1 -1/2 \over N})\mbox{]}
+O({1 \over N^4}). \label{euler}
\end{equation}
The result is 
\begin{eqnarray}
&&{1 \over R}\sum^{\Lambda R-1}_{l=0}
{(l+{1\over 2})/2\pi R \over (l+{1 \over 2})^2 /4\pi^2 R^2 +\alpha^2}
 \nonumber \\ 
&&\!\!=
\int^{\Lambda -{1\over 2R}} _{-{1 \over 2R}}
{(x+1/2R)/2\pi \over (x+1/2R)^2 /4\pi^2  +\alpha^2}dx
+{1\over 48R^2}
\Biggl[ 
{(x+1/2R)^2 /4\pi^2 -\alpha^2 \over ((x+1/2R)^2 4\pi^2 +\alpha^2 )^2 }
\Biggr] 
^{\Lambda -{1 \over 2R}}_{x=-{1 \over2R}} 
+O(1/R^4 ) \nonumber \\ 
&&\!\!=\ln (1+\Lambda ^2 /4\pi^2 \alpha^2 )
+{1 \over 48R^2}{1 \over \alpha^2}
+O(1/R^4 ).
\label{euler2}
\end{eqnarray}
Using this approximation formula, we evaluate the saddle point 
equation (\ref{sadmode}), and find 
\begin{eqnarray}
\lambda_{c}&=&{2\over L}\sinh^{-1}{1\over 2}
+{\pi^3 \over 12}\sinh^{-1}(1/2){L\over R^2}+O(L^3 /R^4) \nonumber \\ 
&=&0.9624/L +2.684{L\over R^2}+O(L^3 /R^4).
\end{eqnarray}
From the above equation, we see the dependence of the correlation 
length on the temperature $T=1/L$ and on the size $R$.
\begin{eqnarray}
\xi^{-1}&=&\sqrt{\lambda_{c}^{2}+1/4R^{2}}, \NN \\
&=&0.9224/L +2.713{L\over R^{2}}+O(L^3 /R^4).
\end{eqnarray}
%\end{enumerate}
%%%%%%%%%%%%%%%%%%%%%%%%%%%%%%%%%%%%%%%%
%%%%Antiperiodic Boundary Condition %%%%
%%%%%%%%%%%%%%%%%%%%%%%%%%%%%%%%%%%%%%%%
%\begin{itemize}
%
\subsection{Case of antiperiodic boundary condition (APBC)}\label{s:APBC} 
%\end{itemize}
%
In the gap equation (\ref{sadmode}), the eigenvalues of the $S^2 $ 
part of $\Delta$ take 
the squares of half odd integers times $R^{-2}$ 
. Those of the $S^1 $ part of 
$\Delta$ also take the square of half odd integers if we 
impose 
the APBC in 
the direction of $S^1$. 
We solve the gap equation (\ref{sadmode}) in three cases 
depending on $R=\infty$, $L=\infty$ and $a\ll L,R$. 
As pointed out in \cite{birrel}, there arise a few subtle problems in
the study of quantum effects in scalar field theories with APBC. 
First, one cannot make use of the
na{\" {\i}}ve effective potential in the computation of the expectation 
value $\sigma$ of 
$\sigma(x)$, since a constant expectation value obviously contradicts
APBC. One can circumvent this difficulty by considering instead linear
modes. The saddle point equation for $\sigma$, then
takes the same form as that in the PBC case, $\sigma\mu=0$. 
Second, the tachyonic modes 
will generally arise after taking account of the quantum effect of
twisted scalar field on untwisted scalar field. 
We will see that the
same type of tachyonic modes appear in the solution of the gap
equation in the APBC case. 
%\renewcommand{\theenumi}{(2\alph{enumi})}
%\begin{enumerate}
%\item 

(2a)\quad $R=\infty $\ \ ($S^1 \times {\bf R}^2 $).

The gap equation can be further reduced  to 
\begin{equation}
{1 \over {\tilde g}}={1 \over {\tilde g_c }}-{1\over 2\pi L}\ln (2\cosh {
L\lambda \over 2}).
\end{equation}
At ${\tilde g}={\tilde g}_c$, this equation has a solution 
$\lambda=\lambda_c =i{\hat \lambda}_{c}$, with 
\begin{equation}
{\hat \lambda}_{c}=(2/L)\cos^{-1}(1/2)=2\pi/3L.
\end{equation} 
This solution is tachyonic, $\lambda_{c}^{2}=\mu=\xi^2 <0$, 
and it should be interpreted that the correct value is 
$\lambda_c =0$. We have not fully understood the mechanism
of dealing with these tachyonic modes.  
While we consider the unstable saddle point where we solve  
the gap equation of this model. 

%\item 
(2b)\quad $L=\infty $\ \ ($S^2 \times {\bf R} $).

The boundary condition becomes immaterial in the limit $L=\infty$. 
Hence we have the same result as (1b). 

%\item 
(2c)\quad General case.

We consider the limit of $R\Lambda,L\Lambda\rightarrow \infty $ 
with $L/R=\delta $ fixed. 
In the case of $\delta\gg 1$, the gap equation can be evaluated  
by means of the Euler-Maclaurin formula  
(\ref{euler}). We have to the order 
of $\delta^{-2}$ as 
\begin{equation}
{1 \over {\tilde g}}={1 \over {\tilde g_c }}-{\lambda\over 16\pi^2}
+O(\delta^{-3}).
\end{equation}
By setting ${\tilde g}={\tilde g_c}$, we 
obtain $\lambda_{c}=0+O(\delta^{-3})$.

In the case of $\delta\ll 1$, we can also make use of 
(\ref{euler}) and find the critical value as 
\begin{equation}
\lambda_c =\lambda_{c}|_{R=\infty}+(\pi^3 /12)(\delta /L)
\end{equation}
formally. 
By the same argument as made in the case (2a), 
the definition should lead $\lambda_c =0$ under the condition 
of at least $\delta\lsim {1\over 4}$. 
(This upper-limit can been seen
if we consider a small $\delta$, which make $\lambda_{c}$ equal to
just 0 not to a pure imaginary number.)
\section{Modular Properties and Behavior of the Specific Heat}
With the above result, let us discuss the physical aspects of the
$O(N)$ NL$\sigma$ models on (semi)compact manifolds.
First we discuss 
the modular invariance of this 
model. The simplest example model with the modular invariance is the two 
dimensional free scalar model on a torus with moduli $\delta$. 
Its free energy $F=T\Sigma_{l=0}^{\infty}\Sigma_{n=0}^{\infty}
\ln(l^2 \delta+n^2 /\delta)$ is invariant under 
$\delta\rightarrow 1/\delta$, if we ignore the regularization of the
divergence. 
We can calculate the free energy of $O(N)$ NL$\sigma$ model on 
$S^{1} \times S^{2}$ with the APBC by means of the saddle point
method. If we choose $g=g_{c}$, the arguments in \ref{s:APBC} enable 
us to calculate free energy as  
\begin{equation}
F=T\sum_{l=0}^{\infty}\sum_{n=0}^{\infty}
(l+1/2)\ln\{ (l+1/2)^2 \delta+(n+1/2)^2 (1/\delta)\} 
\label{free}
\end{equation}
to the order $\delta^{-2}$ in the case of $\delta\gg 1$ and in the
region $\delta\lsim {1\over 4}$ in the case of $\delta\ll 1$.  
Obviously it  is not invariant under $\delta\rightarrow 1/\delta$. 
Hence we deform the definition of the free energy (\ref{free}) as
follows according to \cite{cardy3}.
\begin{equation}
F'=\sum_{l=0}^{\infty}\sum_{n=-\infty}^{\infty}
(l+1/2)(n+1/2)f\Bigl( (l+1/2)^2 \delta+(n+1/2)^2 (1/\delta) \Bigr) , 
\label{dfree}
\end{equation}
where $f$ is a some function. If we exchange the two radii of $S^2 $ and 
$S^1 $, {\it i.e. }$\delta\rightarrow 1/\delta $, the deformed free energy 
defined above is invariant.
Because the every derivative of the free energy 
(\ref{free}) by $\delta$ leads the factor $(n+1/2)^2$, 
the deformed free energy 
(\ref{dfree}) can be regarded as the half derivative by $\delta$ 
of the free energy (\ref{free}).

Second we discuss the attempt of the expansion of Zamolodchikov's
C-theorem in dimensions higher than three.
There exist an very important considerations about the C-theorem
\cite{cappelli} in 
higher-dimensional field theory. The c-functions defined by the finite size 
correction and that introduced by the energy momentum tensor do not 
coincide in higher dimensions in general. We expect it is useful to 
analyze the partition functions to understand the C-theorem. 
We can calculate the specific heat $C(g)={dF(T,g)/dT} ~(,T=1/L)$ of the 
large $N$ $O(N)$ NL$\sigma$ model (PBC) with 
arbitrary $g$ by means of the saddle point method. If we obtain the
mass gap $m=\sqrt{\mu}$ from the gap equation (\ref{saddle}) for the 
coupling constant $g$, the graph of the specific heat 
calculated as 
\begin{equation}
{C(mL)\over N}=\zeta (3)^{-1}\left(Li_{3}(e^{-mL})
+(mL) Li_{2}(e^{-mL})+{(mL)^{3}\over 6} \right)
\end{equation}
is presented as 
\vspace{1cm}
%%%%%%%%%%%%%%%%%%%%%%%%%%%%%%%%
%     c function               %
%%%%%%%%%%%%%%%%%%%%%%%%%%%%%%%%
\begin{center}
Figure 1: Graph of $mL$-$C/N$. 
\end{center}
\vspace{1cm}
We see from the above graph that the specific heat has the local minimum at 
$m\simeq\lambda_{c}$ $(1/\lambda_{c}L=1.039)$, $i.e.$ $g=g_{c}$. 
Therefore, although we consider the model on $S^{1} \times S^{2}$ or 
$S^{1} \times {\bf R}^{2}$, which does not allow the phase transition,
 this behavior of the specific heat  seems to justify to consider the
model on $S^{1} \times S^{2}$ or $S^{1} \times {\bf R}^{2}$ with the
coupling constant $g_{c}$, which is the critical value on ${\bf
R}^{3}$. 
We expect some interpretation in terms of the 
RG. In APBC case, we do not find such a  local minimum of the 
specific heat. Therefore we cannot give a simple justification 
to $g_{c}$ on $S^{1} \times S^{2}$ or $S^{1} \times {\bf R}^{2}$.  

We have a variety of the expansion. Especially, we expect 
the analysis of the four-fermion model or supersymmetric NL$\sigma$ 
model are important because they permit the nonperturbative analysis. 
Furthermore it is reported that the latter has zero $\beta$-function 
independent of the coupling constant \cite{snls}.  And the
combination of the campactification and $\epsilon$-expansion 
will be reported elsewhere \cite{pp1}.

{\bf Figure Caption}
\begin{description}
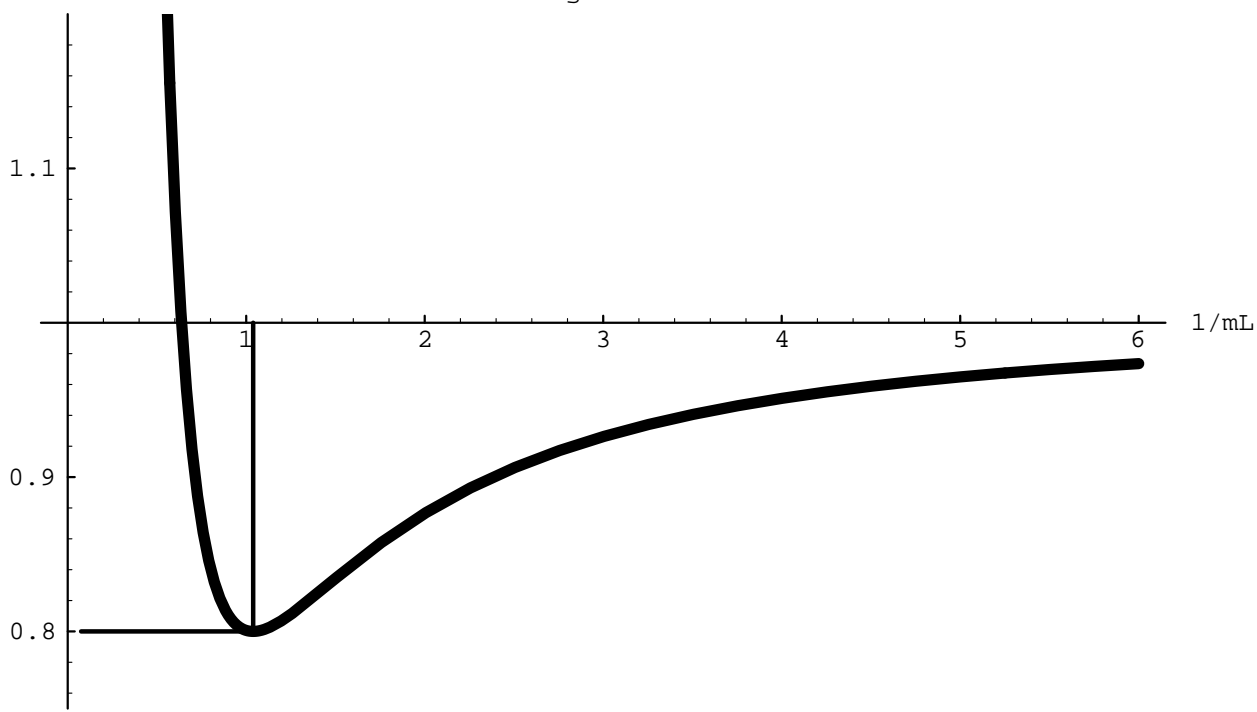
\item[Figure 1.] The graph of the behavior of the specific heat
$C(mL)/N$ in terms of a dimensionless variable $1/mL$.
\end{description}
\end{document}